\documentclass[12pt,letterpaper]{article}

\usepackage{amsmath}
\usepackage{amsfonts}
\usepackage{amssymb}
\usepackage{graphicx}
\def\be{\begin{equation}}
\def\ee{\end{equation}}

\newcommand{\GeV}{{\rm GeV}}

\begin{document}


\begin{flushright} {\footnotesize HUTP-04/A022}  \end{flushright}
\vspace{5mm}
\vspace{0.5cm}
\begin{center}

\def\thefootnote{\fnsymbol{footnote}}

{\Large \bf The shape of non-Gaussianities} \\[1cm]
{\large Daniel Babich$^{\rm a,b}$, Paolo Creminelli$^{\rm a}$ and Matias Zaldarriaga$^{\rm a,b}$}
\\[0.5cm]

{\small 
\textit{$^{\rm a}$ Jefferson Physical Laboratory, \\
Harvard University, Cambridge, MA 02138, USA}} 

\vspace{.2cm}

{\small 
\textit{$^{\rm b}$ Harvard-Smithsonian Center for Astrophysics \\
Cambridge, MA 02138, USA
}}
\end{center}

\vspace{.8cm}

\hrule \vspace{0.3cm} 
{\small  \noindent \textbf{Abstract} \\[0.3cm]
\noindent
We study the dependence on configuration in momentum space of the primordial 3-point
function of density perturbations in several different scenarios: standard slow-roll
inflation, curvaton and variable decay models, ghost inflation, models with higher derivative
operators and the DBI model of inflation. We define a
cosine between the distributions using a measure based on the ability of experiments
to distinguish between them. We find that models fall into two
broad categories with fairly orthogonal distributions. Models where
non-Gaussianity is created at horizon-crossing during inflation and models in which the evolution 
outside the horizon dominates. In the first case the 3-point function is largest for equilateral triangles, 
while in the second the dominant contribution to the signal comes from the influence of long wavelength
modes on small wavelength ones. We show that, because the distributions in
these two cases are so different, translating constraints on parameters
of one model to those of another based on the normalization of the 3-point function for equilateral triangles 
can be very misleading.

\vspace{0.5cm}  \hrule

\def\thefootnote{\arabic{footnote}}
\setcounter{footnote}{0}


\section{Introduction}

Spectacular experimental observations in cosmology caused a certain
optimism about our knowledge of the very early Universe. The results
are sometimes described as a confirmation of the standard slow-roll
inflation paradigm, but this can be rather misleading. What we really
know is that all observations are compatible with a scale invariant
spectrum of adiabatic perturbations with Gaussian statistics and that
these perturbations exist outside the horizon at the time of recombination.  
These facts are too generic to be considered a proof of the standard
picture and in fact non-minimal scenarios or even radically different
proposals are still compatible with the data.

The situation will likely change in the near future. There are three
basic observables which we consider the most relevant both to confirm
or rule out the minimal slow-roll scenario. The experimental limits on
all these parameters are getting close to the interesting range, where
a distinction among different proposals is possible.

{\bf Tilt of the scalar spectrum.} A quite generic prediction of
slow-roll inflation is the deviation from a completely flat spectrum.
This prediction is a consequence of slow-roll itself and it is
therefore rather robust. Although the precise number is
model-dependent, in most models $|n-1|$ is of order $1/N_e$, where
$N_e$ is the number of e-folds to the end of inflation when relevant
scales exit the horizon. Present limits are of order $|n-1| \lesssim
0.05$ ({\em e.g.} \cite{Peiris:2003ff,Tegmark:2003ud,Readhead:2004gy}), 
so that we are entering in the interesting region. 
A deviation from a flat spectrum would strongly support the slow-roll inflation
picture and it would allow to distinguish it from `ghost inflation'
\cite{Arkani-Hamed:2003uz} for example, where $|n-1|$ is expected to
be negligible. However, if no tilt is detected slow-roll inflation
cannot be safely ruled out: it is easy to build models with a tilt as
small as we like.

{\bf Gravity wave (GW) contribution.} The contribution of GWs is
directly related to the value of the Hubble constant $H$ during
inflation. The detection of a GW signal would therefore point towards
models with big vacuum energy ($V^{1/4} \gtrsim 10^{16} \;\GeV$).
Inflationary models fall into two broad categories. Models with small
vacuum energy (which is equivalent to a very small $\epsilon$,
$\epsilon \ll 1/N_e$, as $H/(M_P \sqrt{\epsilon})$ is fixed by the
spectrum normalization) with totally negligible productions of GWs and
models with big vacuum energy (usually with $\epsilon \sim \eta \sim
1/N_e$), where the GW contribution should be close to the present
experimental limit, $r \lesssim 0.5$ ({\em e.g.} \cite{Peiris:2003ff, Tegmark:2003ud}) . The
distinction is quite sharp because the two categories can also be
distinguished by the variation of the inflaton field during inflation:
much smaller than the Planck scale in the first case, comparable to 
the Planck scale in the second. A possible criticism against models with a sensible
production of GWs is that a variation of the inflaton field much
bigger than $M_P$ seems out of control of the effective field theory
\cite{Lyth:1996im}. Extra dimensional UV completions provide
examples in which this is not true \cite{Arkani-Hamed:2003wu}. On the
other hand models with very small $\epsilon$ have been considered
unnatural as they require a hierarchy between the two slow-roll
parameters $\epsilon \ll \eta$ \cite{Khoury:2003vb}. Experiments in
the near future will distinguish between the two possibilities. A
detection of a GW signal would be of great support for the simplest
slow-roll inflation scenario. GWs are in fact usually negligible in
models where additional light fields are responsible for density
perturbations \cite{Lyth:2002my,Dvali:2003em,Dvali:2003ar}, (see however
\cite{Pilo:2004ke}) in the ekpyrotic/cyclic scenario
\cite{Khoury:2004xi} and in ghost inflation
\cite{Arkani-Hamed:2003uz}.

{\bf Non-Gaussianity.} The third observable, which is the main subject
of this paper, is the deviations from a pure Gaussian statistics, {\em
  i.e.} the presence of a 3-point function\footnote{The presence of
  any connected $n$-point function ($n > 2$) indicates a deviation from
  a perfectly Gaussian signal. We concentrate on the 3-point function
  as it is much bigger than the others in all the model we consider.}.
There are two reasons why the study of the 3-point function is
relevant. First of all, in a conventional single field model of
inflation, the 3-point function can be explicitly calculated as a
function of the slow-roll parameters \cite{Maldacena:2002vr,
  Acquaviva:2002ud}. It turns out to be very small: the primordial
fluctuations are Gaussian up to a level of $10^{-6}$ (dimensionless
skewness), which is beyond what we can measure in the near future. Any
deviation from this prediction is therefore a clear sign of departure
from the simplest picture. The 3-point function therefore appears as
the optimal smoking gun for many possible scenarios: additional light
fields besides the inflaton, imprints of heavy physics through higher
dimension operators, ghost inflation, etc.  Conversely, if no significant
level of non-Gaussianity is found, this will favor the simplest
scenario. Also the ekpyrotic and cyclic scenarios, disregarding the open issue
of matching the perturbations across the bounce, can give an extremely 
Gaussian spectrum \cite{Steinhardt}.

Another reason why the detection of the 3-point function would be very
exciting is that it potentially contains a lot of information.  The
3-point function of the curvature perturbation $\zeta$ in momentum
space
\begin{equation}
\label{eq:3point}
\langle \zeta_{\vec k_1}\zeta_{\vec k_2}\zeta_{\vec k_3} \rangle 
\end{equation}
depends on 2 parameters which characterize the shape of the $(\vec
k_1, \vec k_2, \vec k_3)$ triangle, while the dependence under
rescaling of the triangle is fixed by scale invariance\footnote{Given
  the limits we have on the scalar tilt, scale invariance is a very
  good approximation.}. The purpose of this paper is to show that this
function of two parameters contains a lot of information about the
source of non-Gaussianity and that it could be useful to distinguish
among different models. Moreover we will study how the experimental
limits, which are given assuming a particular form of the 3-point
function, change if we modify the shape dependence of
(\ref{eq:3point}).

There are in principle other observables which could turn out to be
relevant, like for example an isocurvature contribution in the
perturbations. Unless a conserved quantity such baryon number prevents it, 
thermal equilibrium is capable of erasing any isocurvature fluctuations imprinted 
early on, so that it is rather difficult to get generic predictions. Therefore 
only for the three observables described above we are confident to enter, with 
the experimental progress in the next few years, in an interesting range. Whatever 
the results turn out to be we will get further insight into the early cosmology.

In section \ref{sec:general} we will describe the general features of the 3-point function
in different models and underline the qualitative differences. In section \ref{sec:3d} we
plot the different functions and we quantify how ``orthogonal'' two distributions are.
In section \ref{sec:2d} we study the effect of projecting the 3d space into a 2d 
Cosmic Microwave Background (CMB) map
and how experimental limits on non-Gaussianity change taking a different shape dependence. 
We leave to the appendix a discussion about the approximation we used in the 2d analysis. 
Conclusions are drawn in section \ref{sec:conclusions}.

\section{\label{sec:general}Shape dependence in different models}

Translational invariance forces the 3-point function (\ref{eq:3point})
to conserve momentum
\begin{equation}
\label{eq:delta}
\langle \zeta_{\vec k_1}\zeta_{\vec k_2}\zeta_{\vec k_3} \rangle = (2\pi)^3 \delta \big(\sum_i \vec k_i\big)
F(\vec k_1, \vec k_2,\vec k_3) \;,
\end{equation}
while scaling invariance implies that the function $F$, symmetric in
its arguments, is a homogeneous function of degree $-6$
\begin{equation}
\label{eq:hom}
F(\lambda \vec k_1,\lambda \vec k_2,\lambda\vec k_3) = \lambda^{-6} F(\vec k_1, \vec k_2,\vec k_3) \;.
\end{equation}
Rotational invariance further reduces the number of independent variables to just 2, for example the
two ratios $k_2/k_1$ and $k_3/k_1$. Note that the function $F$ is real, because the 3-point function in 
position space cannot change if we change sign to all coordinates.

One interesting form for the function $F$ is the one usually assumed
for the analysis of the data (see {\em e.g.} \cite{Komatsu:2003fd}).
The quantity we observe $\zeta$ is not Gaussian but it contains a non-linear ``correction''
\begin{equation}
\label{eq:f_NL}
\zeta(x) =\zeta_g(x) -\frac35 f_{\rm NL}(\zeta_g(x)^2 - \left<\zeta_g^2\right>) \;,
\end{equation}   
where $\zeta_g(x)$ is Gaussian.  Experimental limits are usually put on the scalar variable 
$f_{\rm NL}$ (\footnote{The $3/5$ is introduced so that $f_{\rm NL}$ parametrizes the amplitude of the
  non-Gaussian departures of  matter-era gravitational
  potential on the large scales.}). The most stringent limit comes from the WMAP experiment
\cite{Komatsu:2003fd} 
\begin{equation}
\label{eq:WMAPlimit}
-58 < f_{\rm NL} < 134 \quad {\rm at} \; 95\% \; {\rm C.L.} 
\end{equation}
If we go to Fourier space, eq.~(\ref{eq:f_NL}) implies a function $F$ of the form
\begin{equation}
\label{eq:FFNL}
F_{\rm local}(\vec k_1, \vec k_2,\vec k_3) = 2 (2\pi)^4 (-\frac35 f_{\rm NL} P_{\cal{R}}^2) \cdot
\frac{\sum_i k_i^3}{\prod k_i^3} \;,
\end{equation}
where $P_{\cal{R}}$ is the amplitude of the power spectrum. 
Currently the best constraint on its amplitude comes from the CMB 
anisotropy measurement by the WMAP satellite,  $P_{\cal{R}}^{1/2} \simeq 4.3 \times 10^{-5}$ \cite{Peiris:2003ff}.

Although originally taken as a simple ansatz, this shape dependence
turns out to be physically relevant for many models which predict a
sensible non-Gaussianity. The reason is that eq.~(\ref{eq:f_NL})
describes (at leading order) the most generic form of non-Gaussianity
which is {\em local in real space}. This form is therefore expected
for models where non-linearities develop outside the horizon.  This
happens for all the models in which the fluctuations of an additional
light field, different from the inflaton, contribute to the curvature
perturbations we observe. In this case non-linearities come from the
evolution of this field outside the horizon and from the conversion
mechanism which transforms the fluctuations of this field into density
perturbations. Both these sources of non-linearity give a
non-Gaussianity of the form (\ref{eq:f_NL}) because they occur outside
the horizon. Examples of this general scenario are the curvaton models
\cite{Lyth:2002my}, models with fluctuations in the reheating
efficiency \cite{Dvali:2003em,Dvali:2003ar} and multi-field
inflationary models \cite{Bernardeau:2002jy} (\footnote{In these
  models additional contributions to $F$ not of the local form
  (\ref{eq:FFNL}) can be present; they describe non-Gaussianities
  generated at horizon crossing. Nevertheless the local contribution
  is dominant because it has time to develop outside the horizon for
  many Hubble times before the final conversion to density
  perturbations \cite{Zaldarriaga:2003my}.}).

Being local in position space, eq.~(\ref{eq:FFNL}) describes correlation among Fourier modes of very different
$k$. It is instructive to take the limit in which one of the modes becomes of very long wavelength 
\cite{Maldacena:2002vr}, $k_3 \rightarrow 0$, which implies, due to momentum conservation, that the other 
two $k$'s become equal and opposite.
The long wavelength mode $\zeta_{\vec k_3}$ freezes out much before the others and behaves as a background for 
their evolution.  
In this limit $F_{\rm local}$ is proportional to the power spectrum of the short and long wavelength modes
\begin{equation}
\label{eq:limitloc}
F_{\rm local} \propto \frac1{k_3^3} \frac1{k_1^3} \;.
\end{equation}  
This means that the short wavelength 2-point function $\langle \zeta_{\vec k_1} \zeta_{-\vec k_1} \rangle$ depends 
linearly on the background wave $\zeta_{\vec k_3}$
\begin{equation}
\label{eq:linear}
\langle \zeta_{\vec k_3}\zeta_{\vec k_1}\zeta_{-\vec k_1} \rangle \propto
\langle \zeta_{\vec k_3}\zeta_{-\vec k_3}\rangle \frac\partial{\partial \zeta_{\vec k_3}} \langle 
\zeta_{\vec k_1}\zeta_{-\vec k_1} \rangle \;.  
\end{equation}
From this point of view we expect that any distribution will reduce to the local shape (\ref{eq:FFNL}) in the 
degenerate limit we considered\footnote{The derivative with respect to the background cannot depend
on the relative orientation of $\vec k_1$ and $\vec k_3$, because this would need a derivative 
acting on the background giving a subleading contribution in the limit $\vec k_3 \rightarrow 0$.}, 
if the derivative with respect to the background wave does not vanish.

In standard single field slow-roll inflation the limit $k_3 \rightarrow 0$ is quite easy to predict. As pointed 
out by Maldacena \cite{Maldacena:2002vr}, different points along the background wave are equivalent to shift in time 
along the inflaton trajectory, so that the derivative with respect to the background wave is proportional to the 
tilt of the scalar spectrum. This can be explicitly checked in the full expression of the 3-point function
\cite{Maldacena:2002vr}
\begin{equation}
\label{eq:malda}
F_{\rm stand}(\vec k_1, \vec k_2,\vec k_3) = \frac18 (2\pi)^4 P_{\cal{R}}^2  \cdot \frac{1}{\prod k_i^3}
\left[(3 \epsilon - 2 \eta) \sum_i k_i^3 + \epsilon \sum_{i \neq j} k_i k_j^2 
+8 \epsilon \frac{\sum_{i>j} k_i^2 k_j^2}{k_t}\right] \;,
\end{equation}
where $\epsilon$ and $\eta$ are the usual slow-roll parameters and $k_t \equiv k_1 + k_2 + k_3$.
In the limit $k_3 \rightarrow 0$ eq.~(\ref{eq:malda}) goes as
\begin{equation}
\label{eq:maldalimit}
F_{\rm stand}(\vec k_3 \rightarrow 0) \propto 2 (\eta - 3 \epsilon) \frac1{k_1^3} \frac1{k_2^3} = 
(n_s-1)\frac1{k_1^3} \frac1{k_2^3} \;.
\end{equation}
As expected the tilt in the spectrum $n_s$ fixes the degenerate limit of the 3-point function. Note however 
that expression (\ref{eq:malda}) is {\em not} of the local form (\ref{eq:FFNL}) but contains contributions 
which are important for non-degenerate triangles. If we compare expression (\ref{eq:FFNL}) and (\ref{eq:malda})
and neglect the different shape dependence, we see that standard single-field inflation predicts $f_{\rm NL}$
of order of the slow-roll parameters. 

We have seen that the degenerate limit $k_3 \rightarrow 0$ describes the
effect of a slowly-varying background wave on the 2-point function. In
many models the correlation is much weaker in this limit than in the
local case (\ref{eq:FFNL}). Physically this means that the correlation
is among modes with comparable wavelength which go out of the horizon
nearly at the same time. In this case the 3-point function in the
degenerate limit is suppressed by powers of $k_3$ with respect to the
behaviour of eq.~(\ref{eq:limitloc}). We have correlation among modes
of comparable wavelength in all models in which the non-Gaussianity is
generated by derivative interactions: these interactions become
exponentially irrelevant when the modes go out of the horizon because
both time and spatial derivatives become small, so that all the
correlation is among modes freezing almost at the same time.

One example of this kind of models is obtained if we add higher derivative operators in the usual inflation scenario;
the leading operator of this form is
\begin{equation}
\label{eq:NRop}
\frac1{8 \Lambda^4} (\nabla\phi)^2 (\nabla\phi)^2 \;,
\end{equation}
where $\phi$ is the inflaton.
It is straightforward to calculate the 3-point function after the addition of this operator \cite{Creminelli:2003iq}.
The result is
\begin{align}
\label{eq:myresult}
F_{\rm h.d.}(\vec k_1, \vec k_2,\vec k_3) = & \frac18 (2\pi)^4 P_{\cal{R}}^2 \frac{\dot\phi^2}{\Lambda^4} \cdot  \nonumber \\ &
\cdot \frac1{\prod k_i^3}  \left[\frac1{k_t^2} \Big(\sum_i k_i^5 + \sum_{i \neq j} (2 k_i^4 k_j
- 3 k_i^3 k_j^2) + \sum_{i \neq j \neq l} (k_i^3 k_j k_l -4 k_i^2 k_j^2 k_l) \Big) \right]\;.
\end{align}
The ratio $\dot\phi^2/\Lambda^4$, where $\dot\phi$ is the velocity of
the inflaton, is expected to be less than one in the regime in which
we can trust an effective field theory description with cut-off
$\Lambda$ and neglect the infinite set of higher dimension operators.
Therefore the effect cannot be too big: comparing the previous
expression with eq.~(\ref{eq:FFNL}) and neglecting the shape
dependence we expect roughly $f_{\rm NL} \lesssim 1$, unless we want
to enter into the regime where higher order corrections are
unsuppressed.  It is easy to check that the expression in brackets in
eq.~(\ref{eq:myresult}) vanishes as $k_3^2$ in the limit $k_3
\rightarrow 0$ \cite{Creminelli:2003iq}. The correlation is therefore
highly suppressed in the degenerate limit with respect to the local
shape (\ref{eq:FFNL}): the additional powers of $k_3$ come from the
derivatives in the operator (\ref{eq:NRop}) acting on the background
wave. The correlation is among modes of comparable wavelength, because
the higher derivative interaction vanishes exponentially outside the
horizon. 

A model of inflation based on the Dirac-Born-Infeld (DBI) action has recently been proposed \cite{Silverstein:2003hf,Alishahiha:2004eh}. 
This model predicts significant non-Gaussianities and gravity waves. The predicted form of the 3-point function is the same as in
equation (\ref{eq:myresult}), but the level of non-Gaussianity is much bigger because higher derivatives terms are crucial for
the inflaton dynamics.

As a final example we consider the 3-point function predicted by `ghost inflation'. Without entering into the details
of the model \cite{Arkani-Hamed:2003uz}, we stress that also in this case the 3-point function is generated by a 
derivative interaction, so that we expect the same qualitative behaviour than in the previous example, with a substantial 
level of non-Gaussianity. The explicit form of the 3-point function is not
illuminating
\begin{align}
\label{eq:ghost}  F_{\rm ghost}(\vec k_1, \vec k_2,\vec k_3) =  
& 2 \sqrt{2} \;\pi^{31/10} \;\Gamma\big(\frac14\big)^{1/5} \; \beta \; \alpha^{-8/5} P_{\cal{R}}^{8/5} \cdot  
\\ \nonumber & \cdot \frac1{\prod_i k_i^3} 2 \,{\rm Re}\,
\int_{-\infty}^0 d\eta \;\eta^{-1} F^*(\eta)
F^*\left(\frac{k_2}{k_1} \eta\right) F^{\prime
*}\left(\frac{k_3}{k_1}\eta\right) k_3 (\vec k_1 \cdot \vec k_2) +
{\rm symm.}
\end{align} 
where the contour of integration is $\propto (-1-i)$, $\alpha$ and $\beta$ are unknown order one coefficients and 
\begin{equation} 
\label{eq:F} 
F(x) = \sqrt{\frac{\pi}{8}} (-x)^{3/2} H^{(1)}_{3/4}(x^2/2) \;.
\end{equation}
It can be checked that in the limit $k_3 \rightarrow 0$ the integral goes like $k_3^2$. As in the previous example
the correlation is therefore suppressed for modes with very different wavelength.

As the explicit expressions of the 3-point functions we showed have progressively become more and more complicated, in the
next section we show the explicit plots of the functions, so that their behaviours and differences can be better 
appreciated.

\section{\label{sec:3d}A 3D comparison}

Imagine that we measure the density perturbation $\zeta_{\vec k_i}$ in a 3-dimensional survey. We assume that the 3-point function
is of the form
\begin{equation}
\label{eq:shape}
\langle \zeta_{\vec k_1}\zeta_{\vec k_2}\zeta_{\vec k_3} \rangle = A \cdot (2\pi)^3 \delta \big(\sum_i \vec k_i\big)
F(\vec k_1, \vec k_2,\vec k_3) 
\end{equation}
and we want to use the data to measure the overall amplitude $A$ (\footnote{We should choose some standard normalization for $F$ to give sense to the overall
amplitude $A$. We will come back to this point later.}). It is easy to check that the best estimator for $A$, in the limit of 
small non-Gaussianity, is
\begin{equation}
\label{eq:estim}
\hat A = \frac{\sum_{\vec k_i} \zeta_{\vec k_1} \zeta_{\vec k_2} \zeta_{\vec k_3} F(\vec k_1, \vec k_2,\vec k_3)/(\sigma^2_{k_1}\sigma^2_{k_2}\sigma^2_{k_3})}{
\sum_{\vec k_i} F(\vec k_1, \vec k_2,\vec k_3)^2/(\sigma^2_{k_1}\sigma^2_{k_2}\sigma^2_{k_3})} \;,
\end{equation}
where $\sigma^2_{k_i}$ is the variance of a given mode and the sums run over all triangles in momentum space. This is the estimator with the least variance.  
Expression (\ref{eq:estim}) naturally defines a scalar product between two distributions $F_1$ and $F_2$
\begin{equation}
\label{eq:prod}
F_1 \cdot F_2 \equiv \sum_{\vec k_i}  F_1(\vec k_1, \vec k_2,\vec k_3) F_2(\vec k_1, \vec k_2,\vec k_3)/(\sigma^2_{k_1}\sigma^2_{k_2}\sigma^2_{k_3}) \;.
\end{equation}
Its intuitive meaning is clear: if two distributions have a small scalar product, the optimal estimator (\ref{eq:estim}) for one 
distribution will be very bad in detecting non-Gaussianities with the other shape and vice versa. We will be more quantitative below. But first 
of all we want to use this scalar product to make meaningful plots of the different shapes we described in the previous section.

As we discussed the function $F$ depends on only two independent variables. We choose them to be $x_3 \equiv k_3/k_1$ and $x_2 \equiv k_2/k_1$ and we further assume   
$x_3 \leq x_2$ to avoid considering the same configuration twice. The inequality $x_3 \geq 1-x_2$ follows from the triangular inequality.
Looking at eq.~(\ref{eq:prod}) we see that in the definition of the scalar product there is a factor $x_2^3 x_3^3$ coming from the two spectra in the 
denominator which are approximately scale-invariant. Furthermore a measure $x_2 x_3$ is required to go from the 3D sum over modes to the integral over 
$x_2$ and $x_3$. We conclude that the most meaningful quantity to plot is 
\begin{equation}
\label{eq:plot}
F(1,x_2,x_3) \;  x_2^2 x_3^2 \;,
\end{equation}
so that the integral of the product of two functions we plot gives directly the scalar product.

\begin{figure}[!ht]             
\begin{center}
\includegraphics[width=12cm]{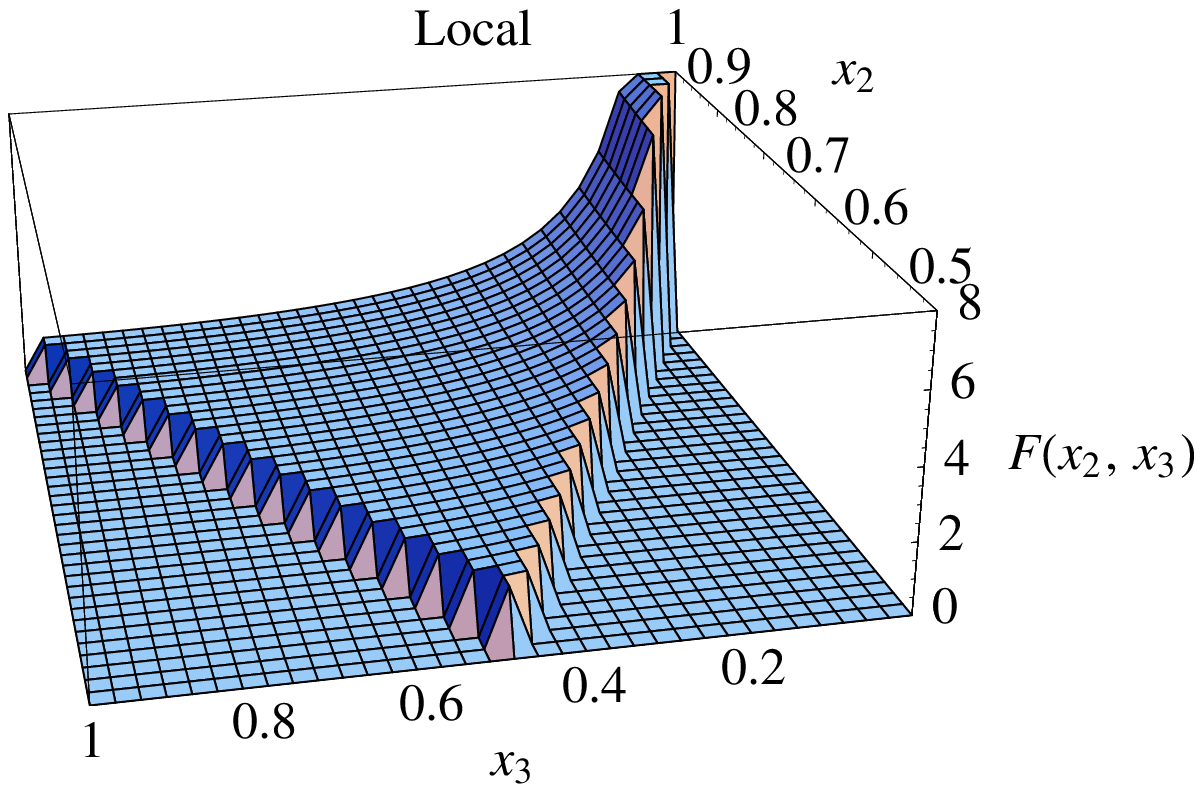}
\caption{\label{fig:local} \small Plot of the function $F(1,x_2,x_3) \;  x_2^2 x_3^2$ for the local distribution (\ref{eq:FFNL}). The figure is normalized to have
value 1 for equilateral configurations $x_2 = x_3 =1$ and set to zero outside the region $1-x_2 \leq x_3 \leq x_2$.}
\vspace{1cm}
\includegraphics[width=12cm]{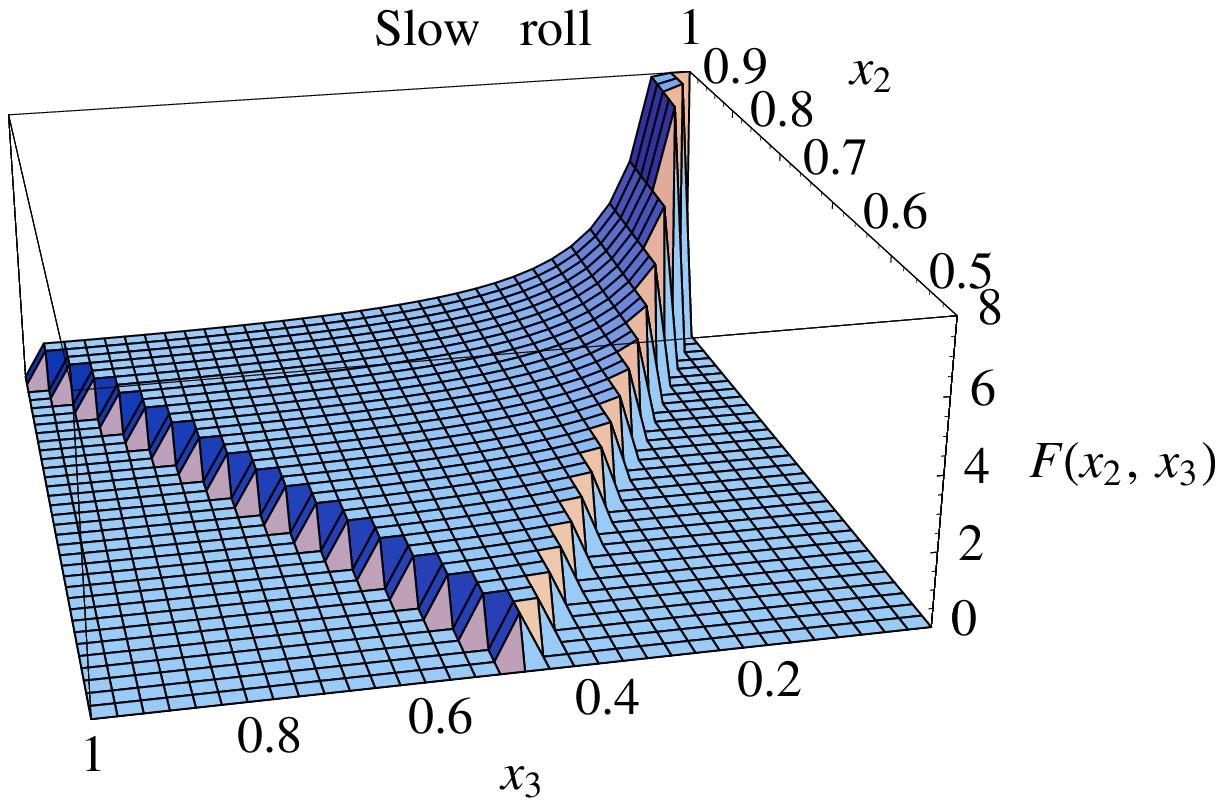}
\caption{\label{fig:malda} \small Plot of the function $F(1,x_2,x_3) \;  x_2^2 x_3^2$ for the usual slow-roll inflation (\ref{eq:malda}) with $\epsilon = \eta = 1/30$. 
The figure is normalized to have value 1 for equilateral configurations $x_2 = x_3 =1$ and set to zero outside the region $1-x_2 \leq x_3 \leq x_2$.}
\end{center}
\end{figure}

\begin{figure}[ht!]             
\begin{center}
\includegraphics[width=12cm]{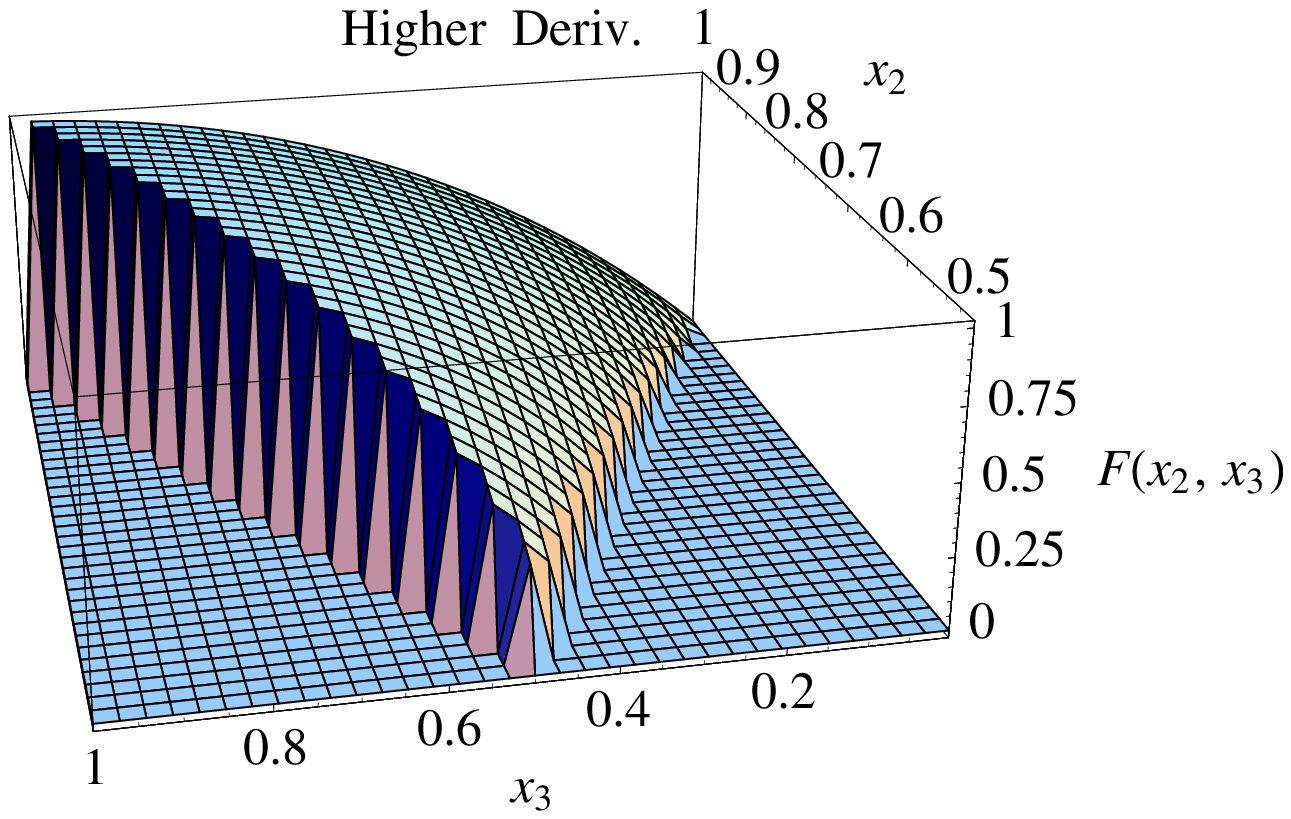}
\caption{\label{fig:hd} \small Plot of the function $F(1,x_2,x_3) \;  x_2^2 x_3^2$ for non-Gaussianities generated by higher derivative interactions 
(\ref{eq:myresult}) and in the DBI model of inflation \cite{Silverstein:2003hf,Alishahiha:2004eh}. 
The figure is normalized to have value 1 for equilateral configurations $x_2 = x_3 =1$ and set to zero outside the region $1-x_2 \leq x_3 \leq x_2$.}
\vspace{1cm}
\includegraphics[width=12cm]{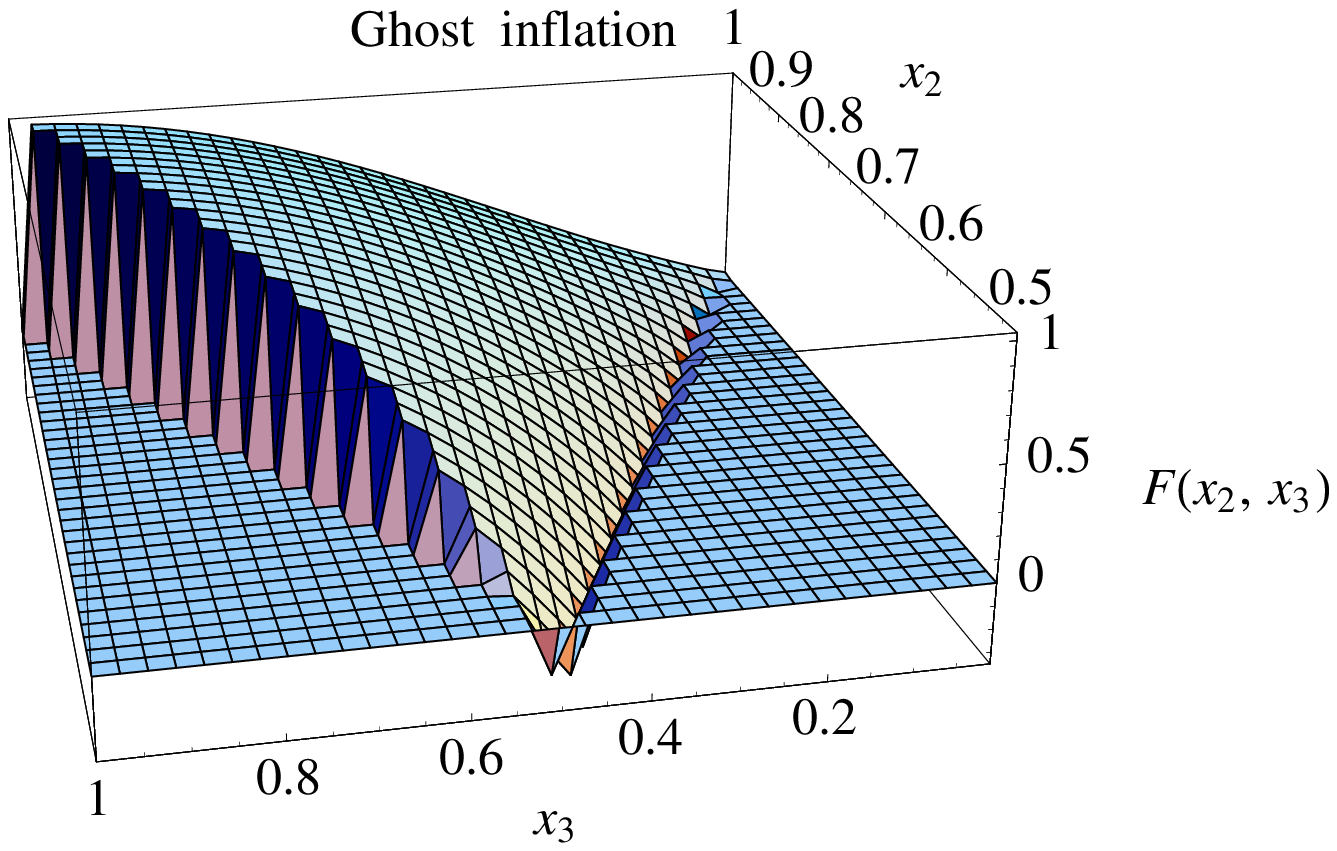}
\caption{\label{fig:ghost} \small Plot of the function $F(1,x_2,x_3) \;  x_2^2 x_3^2$ for ghost inflation (\ref{eq:ghost}). 
The figure is normalized to have value 1 for equilateral configurations $x_2 = x_3 =1$ and set to zero outside the region $1-x_2 \leq x_3 \leq x_2$.}
\end{center}
\end{figure}

In figure \ref{fig:local}, \ref{fig:malda}, \ref{fig:hd} and
\ref{fig:ghost} we show the shape dependences discussed in the
previous section. To avoid showing equivalent configurations twice,
the function is set to zero outside the triangular region $1-x_2 \leq x_3 \leq x_2$. 
In the first two figures we see, as expected, that the
``signal'' is concentrated on degenerate triangles $x_3 \simeq 0$,
$x_2 \simeq 1$, while in the same configuration the third and fourth
plots are suppressed. In these two cases the correlation is bigger
among modes of comparable wavelength, {\em i.e.} equilateral
configurations $x_2 \simeq x_3 \simeq 1$.

We want to be quantitative about the shape difference of the distributions. From the scalar product (\ref{eq:prod}) we can easily define the cosine 
between two distributions
\begin{equation}
\label{eq:cos}
\cos(F_1,F_2) = \frac{F_1 \cdot F_2}{(F_1 \cdot F_1)^{1/2} (F_2 \cdot F_2)^{1/2}} \;,
\end{equation}
which will be a number between $-1$ and 1 (\footnote{Obviously the sign is irrelevant as it can be switched 
by changing the sign of one of the $F$'s.}), 
which tells us how orthogonal two shapes are. If the cosine deviates sensibly from 1, the distinction between two shapes is easy, assuming that a
3-point function has been detected. We numerically calculated the cosine between the 3-point 
functions we discussed with respect to the local distribution, usually assumed in the data analysis. The results are given in table \ref{results}. 
We see, as expected, that the distributions given by higher derivative terms and ghost inflation are not ``collinear'' with the local
distribution: the cosine deviates significantly from 1. The distribution predicted by the conventional slow-roll scenario is on the other hand quite close to the 
local distribution, unless $n-1= 2 (\eta-3 \epsilon)=0$ in which case the spectrum is scale invariant and the 3-point function looks quite similar to 
models with derivative interactions. In going from positive to negative tilt the 3-point function changes sign and close to the transition the
cosine with the local model is close to zero.

There is another interesting quantity we can calculate to compare
different distributions. For the local distribution we can take $A =
f_{\rm NL}$ in eq.~(\ref{eq:shape}) and normalize all the other
distributions at the equilateral configuration. For every distribution
we will have an overall amplitude $f_{\rm NL}^{\rm equil.}$, which can
be directly compared to the local case for an equilateral
configuration. Imagine now that a 3-dimensional set of data is used to
get a limit on $f_{\rm NL}$ for the local distribution. How can we
translate this into a limit for $f_{\rm NL}^{\rm equil.}$ for another
distribution? We can define a ``fudge factor'' which converts limit
from $f_{\rm NL}$ to $f_{\rm NL}^{\rm equil.}$ for another shape
dependence. From eqs. (\ref{eq:estim}) and (\ref{eq:prod}) we easily
obtain that the fudge factor $f$ is 
\begin{equation}
\label{eq:3dfudge}
f(F) \equiv \frac{F \cdot F_{\rm local}}{F_{\rm local} \cdot F_{\rm local}} \;. 
\end{equation}
The limit on $f_{\rm NL}^{\rm equil.}$ of a given distribution will be
the usual $f_{\rm NL}$ parameter divided by $f(F)$.  Obviously
this procedure is not optimal, because we are using an estimator
(\ref{eq:estim}) appropriate for the local distribution to set limits
on a different angular dependence. Anyway it is an easy and fast way
to get approximate limits without doing the full analysis with a new
shape dependence. The fudge factors for the different distributions are
given in table \ref{results}. We see that the fudge factor is much
smaller than 1 for the distribution generated by higher derivative
terms and for ghost inflation.

It is interesting to rewrite the definition of $f(F)$ as
\begin{equation}
\label{eq:3dfudge2}
f(F) = \frac{F \cdot F_{\rm local}}{F_{\rm local} \cdot F_{\rm local}} = \cos(F, F_{\rm local}) \left(\frac{F \cdot F}{F_{\rm local} 
\cdot F_{\rm local}}\right)^{1/2}\;. 
\end{equation}
We see that the fudge factor is proportional to the cosine between the
distributions. This suppression can be eliminated using an optimal
estimator for the distribution of interest. The other factor is the
ratio between the two norms ({\em i.e.} the overall signal) once the
functions are normalized at the same value for equilateral triangles;
this cannot be changed by the analysis.  Obviously for the
distributions of fig.~(\ref{fig:hd}) and (\ref{fig:ghost}) this ratio
is quite suppressed as evident from the plots. That explains the
smallness of the fudge factors for these two distributions. For example for the ghost model the suppression is a factor of 16, 
where approximately a factor of 3 comes from the cosine and a factor of 5.5 from the ratio of norms.

 \begin{table}[t]
 \begin{center}
 \begin{tabular}{|c|c|c|c|c|}
        \hline 
        Distribution & 3d Cosine & 3d Factor & 2d Cosine & 2d Factor \\
        \hline 
        Ghost  & 0.33 & 0.06 & 0.52 & 0.16 \\
        Higher Deriv. & 0.45 & 0.10 &  0.64 & 0.24 \\
        \hline
        $\epsilon = 1/30 $, $\eta = 1/30$ & 0.99 & 0.73 & 0.99 & 0.76 \\
	$\epsilon = 1/300 $, $\eta = 1/30$ & 1.00 & 1.12 & 1.00 & 1.11 \\
        \hline
  \end{tabular}
  \end{center}
  \caption{\label{results} \small The 3d and 2d Cosines and Fudge Factors for
    several different primordial distributions. The values for the
    DBI model coincide with those of the higher derivative distribution. In the limit $\epsilon \rightarrow 0$
    the standard inflation distribution reduces to the local distribution: cosines and fudge factors goes to 1. 
    The 2D numbers are obtained using instrumental noise and band width appropriate for WMAP.}
  \end{table}

Finally we want to mention the fact that non-linear evolution of modes
inside the horizon also creates non-Gaussianity in the density field. It
can be observed for example in the local distribution of
galaxies (see \cite{Bernardeau:2001qr} for a review).  These
non-Gaussianities are not scale invariant, a fact that can be used to
separate them from the primordial contributions
\cite{Scoccimarro:2003wn}.  As an example in figure \ref{fig:nl} we
show the shape of the three point function of the density for
$k=0.1\ h\ {\rm Mpc}^{-1}$ for a qualitative comparison with our
previous figures of primordial non-Gaussianities. These
non-Gaussianities are not only scale dependent,
but they also have a different dependence on the triangle shape. The
density 3-point function peaks for collinear configurations, that
is when the three wavevectors are parallel. This happens because
gravity generates density and velocity-divergence gradients that are
parallel to the velocity flows \cite{Scoccimarro:1996jy}.

\begin{figure}[t]             
\begin{center}
\includegraphics[width=12cm]{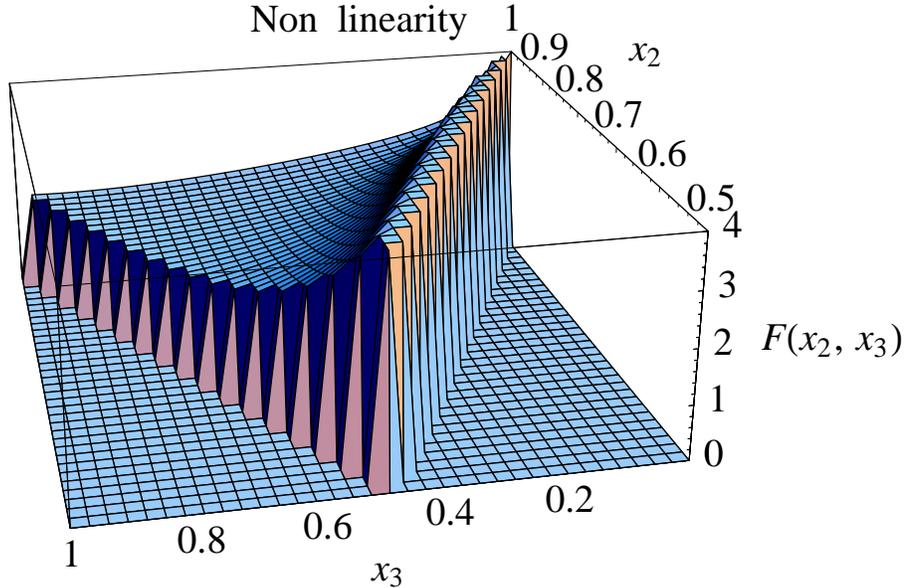}
\caption{\label{fig:nl} \small Shape dependence of the density 3-point
  function for $k=0.1\ h\ {\rm Mpc}^{-1}$. The signal is largest for
  collinear triangles, when the three wavevectors are parallel. These
  fluctuations are not scale invariant so both the amplitude and
  details of the shape are functions of the scale. We present results for
  a particular wavelength for illustrative purposes.}
\end{center}
\end{figure}

\section{\label{sec:2d}A 2D comparison} 

{\bf Non-Gaussianity in the CMB.} The initial curvature perturbations
produced during inflation cause corresponding fluctuations in the matter
species in the universe. These fluctuations, after being modified by
gravitational and hydrodynamical evolution, produce anisotropies in
the CMB. Therefore non-Gaussian statistics of the CMB can be used to
constrain the non-Gaussian statistics of the underlying perturbations.
Most of the fluctuations that we observe in the CMB were imprinted at
the epoch of last scattering, around a redshift of $z\sim 1100$.
Since the fluctuations were very small at that time it is possible to
calculate the radiative transfer of the CMB using linear theory. In
this regime any non-Gaussianity in the CMB will be directly related to
primordial non-Gaussianity. In reality there are non-linear
corrections to the gravitational and hydrodynamical evolution which
will produce non-Gaussianities even if the primordial perturbations are
Gaussian.  One expects such corrections, being of second order in the
perturbation, to produce an equivalent $f_{\rm NL}\sim 1$. This is
below the current experimental limit but non-linear effects might
become important for future experiments. We will ignore them in what
follows.

The correspondence between the non-Gaussianities in the CMB sky and
those of the primordial curvature is not direct because of the
gravitational and hydrodynamical evolution before the epoch of last
scattering. Moreover with the CMB one maps a 2-d surface of the
universe. The temperature on this surface is a projection of the 3-d
curvature perturbations in the surface's vicinity. This fact
introduces further complications if we are interested in the $\vec{k}$ dependence of 
the primordial 3-point function, not just in its amplitude. In a CMB map one can measure only the
components of $\vec{k}$ that are parallel to the plane of the sky, but
not the perpendicular one. As a result a measurement of the 3-point
function of the CMB temperature for a triangle of one particular
shape, will receive contributions from 3-d triangles with a variety of
shapes. Moreover the perturbations in the CMB are no longer scale invariant. The evolution inside the horizon 
imprints several scales in the spectrum, like the scale of the sound horizon or the scale of photon diffusion. 
These departures from scale invariance, though mild, complicate the analysis. 

{\bf CMB statistics.} As for the curvature fluctuations in 3-d, we
will study the 3-point function of the CMB temperature in Fourier
space. We will follow the notation of \cite{Komatsu:2002db}. The
temperature fluctuations on the sky are expanded in spherical
harmonics,
\begin{equation}
\frac{\Delta T}{T}(\hat n) = \sum_{lm} a_{lm} Y_{lm}(\hat n) \;.
\end{equation}
We consider the 3-point function $\langle a_{l_1 m_1} a_{l_2 m_2} a_{l_3 m_3} \rangle$ and,
assuming rotational invariance, construct the angular averaged bispectrum 
 \begin{equation}
     B(l_1,l_2,l_3) = \sum_{m_1,m_2,m_3} \left(\begin{array}{ccc} l_1 & l_2 & l_3 \\ m_1 & m_2 & m_3 \end{array}\right) 
     \langle a_{l_1 m_1} a_{l_2 m_2} a_{l_3 m_3} \rangle \;. 
   \end{equation}

As we did for the curvature perturbations, we can now define a scalar product in 2-d
\begin{equation}
\label{dot}
        B_1\cdot B_2 = \sum_{l_1,l_2,l_3} 
          B_1(l_1,l_2,l_3) B_2(l_1,l_2,l_3)/(f_{l_1,l_2,l_3}C_{l_1} C_{l_2} C_{l_3}) \;,
\end{equation}
where $f_{l_1,l_2,l_3}$ is a combinatorial factor equal to 1 if the
three $l$'s are different, to 2  if two of them are equal and to 6 if
all of them are equal.  The noise  in the
denominator of (\ref{dot}) has been calculated in the Gaussian limit
and includes instrument noise and beam width in the standard way
\cite{Komatsu:2002db,Knox}.  The dot product can be used to define a
2-d cosine, 
\begin{equation}
\label{eq:2dcos}
\cos(B_1,B_2) \equiv \frac{B_1 \cdot B_2}{(B_1 \cdot B_1)^{1/2} (B_2 \cdot B_2)^{1/2}} \;,
\end{equation}
and a 2-d  fudge factor,
\begin{equation}
\label{eq:2dfudge}
f(B) \equiv \frac{B \cdot B_{\rm local}}{B_{\rm local} \cdot B_{\rm local}} \;. 
\end{equation}

{\bf Calculation of CMB Bispectra.} The temperature anisotropies on the sky are linearly related to the underlying curvature perturbations. 
The contribution to the temperature fluctuations at multipole $l$ from a curvature fluctuation with wavenumber $k$ is encoded in the radiation 
transfer function $\Delta^T_l(k)$. In particular we have, 
 \begin{eqnarray}\label{bispect}
      \langle a_{l_1 m_1} a_{l_2 m_2} a_{l_3 m_3} \rangle  = 
      (4\pi)^3 i^{l_1+l_2+l_3}\int \frac{d^3\vec{k}_1}{(2\pi)^3} \frac{d^3\vec{k}_2}{(2\pi)^3} \frac{d^3\vec{k}_3}{(2\pi)^3} 
      Y^*_{l_1 m_1}(\hat{k}_1) Y^*_{l_2 m_2}(\hat{k}_2) Y^*_{l_3 m_3}(\hat{k}_3) \\
      \times (2\pi)^3 \delta^{(3)}(\sum_i \vec k_i) F(k_1,k_2,k_3) \Delta^T_{l_1}(k_1)
      \Delta^T_{l_2}(k_2)\Delta^T_{l_3}(k_3) \;. \nonumber
\end{eqnarray}
The radiation transfer function $\Delta^T_l(k)$ can be calculated with publicly available software such as CMBFAST \cite{Seljak:1996is}. Expressing the
   $\delta$ function as an exponential and  expanding it in spherical harmonics and  Bessel 
   functions we get,
    \begin{eqnarray}\label{bispect2}
          \langle a_{l_1 m_1} a_{l_2 m_2} a_{l_3 m_3} \rangle = 
          \int \frac{2 k^2_1 dk_1}{\pi} \frac{2 k^2_2 dk_2}{\pi} \frac{2 k^2_3 dk_3}{\pi}  
         \int d^2\hat{x} Y^*_{l_1 m_1}(\hat{x}) Y^*_{l_2 m_2}(\hat{x})Y^*_{l_3 m_3}(\hat{x}) 
         \\ \times \int\limits^{\infty}_0 x^2 dx \; j_{l_1}(k_1x) j_{l_2}(k_2x) j_{l_3}(k_3x)
        F(k_1,k_2,k_3) \Delta^T_{l_1}(k_1) \Delta^T_{l_2}(k_2) \Delta^T_{l_3}(k_3) \;. \nonumber 
   \end{eqnarray}

In three dimensions, translation invariance forced the three $\vec{k}$'s in the 3-point function to add to zero. In 2-d the equivalent constraint of rotational 
invariance in enforced by the  Gaunt integral
        \begin{eqnarray}\label{gaunt}
        \mathcal{G}^{m_1,m_2,m_3}_{l_1,l_2,l_3} &=& \int d^2\hat{x} Y_{l_1 m_1}(\hat{x}) Y_{l_2 m_2}(\hat{x})Y_{l_3 m_3}(\hat{x}) \nonumber \\
           &=& \sqrt{\frac{(2l_1+1)(2l_2+1)(2l_3+1)}{4\pi}} 
             \left(\begin{array}{ccc} l_1 & l_2 & l_3 \\ 0 & 0 & 0 \end{array}\right)
             \left(\begin{array}{ccc} l_1 & l_2 & l_3 \\ m_1 & m_2 & m_3 \end{array}\right) \;.
        \end{eqnarray}
It forces $l_1$, $l_2$ and $l_3$ to satisfy a triangle inequality.  The integral
   \begin{equation}\label{besint}
       \mathcal{C}_{l_1,l_2,l_3}(k_1,k_2,k_3) = \int \limits^{\infty}_0 x^2 dx \; j_{l_1}(k_1x) j_{l_2}(k_2x) j_{l_3}(k_3x) \;,
   \end{equation}
   determines the strength with which a 3d triangle contributes to a 2d triangle. When considering the
   2-point function, the equivalent of eq.~(\ref{gaunt}) is $\delta_{l_1,l_2}\delta_{m_1,m_2}$ and
   eq.~(\ref{besint}) becomes proportional to $\delta^{(1)}(k_1-k_2)$. 
Using these definitions we can write the reduced CMB bispectrum in a more convenient form
\begin{eqnarray}\label{bispect3}
B(l_1,l_2,l_3) & & = \sqrt{\frac{(2l_1+1)(2l_2+1)(2l_3+1)}{4\pi}} 
\left(\begin{array}{ccc} l_1 & l_2 & l_3 \\ 0 & 0 & 0 \end{array}\right) \\
& & \!\!\!\!\!\!\!\times \int \frac{2 k^2_1 dk_1}{\pi} \frac{2 k^2_2 dk_2}{\pi} 
\frac{2 k^2_3 dk_3}{\pi}  
\mathcal{C}_{l_1,l_2,l_3}(k_1,k_2,k_3) F(k_1,k_2,k_3) \Delta^T_{l_1}(k_1) \nonumber
\Delta^T_{l_2}(k_2)\Delta^T_{l_3}(k_3) \;. 
   \end{eqnarray}

{\bf Numerical Challenges.} The evaluation of equation
(\ref{bispect3}) is numerically very challenging. It not only involves 
a four dimensional integral, but both the Bessel functions in eq.~(\ref{besint}) and the
radiation transfer functions in eq.~(\ref{bispect3}) are very
rapidly oscillating. For example the radiation transfers function
oscillate when $k$ changes by order the inverse of the distance to
the last scattering surface, $\Delta k\sim 1/d_{\rm LSS}$. Moreover the
contribution to multipole $l$ comes preferentially from modes with
wavenumber $k\sim l/d_{\rm LSS}$. Thus for $l \sim 1000$ the radiation
transfer functions have many oscillations in the $k$ range of
interest.  We can estimate that in order to even crudely calculate
the integral in equation (\ref{bispect3}) the transfer function
needs to be evaluated in $200$ values of $k$. Since there are three
integrals over $k$ the required number of evaluations is roughly
$(200)^3 \sim 10^7$.  In addition there is the integral over the
three spherical Bessel function, eq.~(\ref{besint}), which is
extremely difficult to evaluate numerically due to the oscillatory
nature of its integrand.  Let us assume that it can be done with $10^3$ operations.  
We need to calculate these integrals roughly $l^3_{max}$ times because of the sum which appears 
in the dot product (\ref{eq:2dcos}). For $l_{max} \approx 1000$ this is $10^9$. We must
perform roughly $10^{19}$ operations to compute the dot products by brute force.

It is clear that an alternative way to evaluate these integrals must be
developed. For the particular case of the local distribution, $F_{\rm local}$ in 
eq.~(\ref{eq:FFNL}) can be expressed as a product of functions of $k_1$, $k_2$ and $k_3$ separately and 
the integral in equation (\ref{bispect3}) can be split and done more easily \cite{Komatsu:2002db}. 
We cannot do this in general because some of the distributions we are considering depend
of $k_t$ and cannot be factorized.

In the appendix we present a technique to evaluate equation
(\ref{bispect3}) using the flat sky approximation. Under this
assumption a simple change of variables can eliminate most of the
oscillations in the transfer functions and the integral can then be
evaluated numerically with little effort.

{\bf Results.} The results of cosines and fudge factors with respect to the local distribution are listed in Table \ref{results}. 
We used the instrumental noise and beam width appropriate for the WMAP experiment \cite{Bennett:2003bz}, even though we do not expect big differences
for other experiments. The first two entries are for ghost inflation and for the 3-point function generated by higher derivative operators 
(which coincides with the one in the DBI model). 
We also show the results for the standard slow-roll inflation distribution calculated for different values of the slow-roll parameters. 

Table \ref{results} shows that the cosines between distributions calculated for an ideal 3-D  experiment and those calculated for a CMB map are 
quite similar. That is to say, models that are distinguishable in 3-D are also distinguishable in a 2-D survey. 
For a given 2-D Fourier mode, the components parallel to the plane of the sky of the 3-D modes that contribute to it are fixed, 
$\vec{k}^{\parallel} = \vec{l}/ d_{\rm LSS}$. However Fourier modes with all possible values of the wavevector component perpendicular to the plane 
of the sky $k_\perp$ can contribute. As a result one expects that the configurational dependence of the 3-point function in 2-D be somewhat washed out 
relative to the 3-D case. Triangles with all different shapes in 3-D contribute to a given triangle in 2-D. This effect however is rather mild as 
evident in Table \ref{results}.  
The fact that the primordial bispectrum is scale invariant, {\it i.e.} its amplitude is proportional to $k^{-6}$, implies that the dominant contribution 
comes from modes with $k_\perp \sim 0$; the information on the shape is conserved. In a sense this is the same reason why we see acoustic peaks in the 
power spectrum. In that case one could also argue that modes with different wavenumbers $k$ contribute to any given $l$ and thus the acoustic oscillations 
in the 3-D transfer function would be washed out in the temperature power spectrum. Clearly that only happens to a small degree. 
 
From the table we can infer how the constraints on $f_{\rm NL}$ from
the WMAP 1-yr data convert to limits on $f_{\rm NL}^{\rm equil.}$ for
different distributions. For example the allowed interval 
\begin{equation} -58 <
f_{\rm NL} < 134 \quad {\rm at} \; 95\% \; {\rm C.L.}  
\end{equation} 
is
approximately degraded to 
\begin{equation} -360 < f_{\rm NL}^{\rm equil.} < 840
\quad {\rm at} \; 95\% \; {\rm C.L.}  
\end{equation} 
for the ghost model. The suppression is roughly a factor of 6. A factor of 3 comes from the difference in norms ({\em i.e.} the overall signal once
the functions are normalized at the equilateral configuration) and a factor of 2 from the cosine. This last piece could be eliminated by optimizing 
the data analysis.

\section{\label{sec:conclusions}Conclusions}

Deviations from Gaussianity could become a very important probe of
the early universe physics, responsible for the density inhomogeneities
we observe in the universe today. The minimal slow-roll model of
inflation predicts negligible non-Gaussianities (at the level of
$10^{-6}$) but many of the alternatives predict levels substantially
larger.

The discovery of a 3-point function could provide
substantial additional information on the mechanisms that generated
the non-Gaussianities through its dependence on triangle shape. In
this paper we studied that dependence for some of the best motivated
alternatives and quantified the observability of the differences in
shapes. We concluded that there are broadly two classes of shapes for
the 3-point function. In models such as ghost inflation,
the DBI model or when there is significant imprint from heavy
physics through higher derivative operators, the non-Gaussianity peaks
for ``equilateral-type'' configurations. For models where the
non-Gaussianities are produced outside the horizon the shape of the
3-point function peaks in the collapsed triangle limit, when the
wavelength of one of the modes is much larger than that of the other
two. This limit is well described by the local model.

We quantified these differences by introducing a cosine between the
distributions with a measure based on the signal to noise, that is the
ability of experiments to distinguish the different shapes. We
found that the cosines between distributions are very similar for CMB experiments and ideal
3D experiments, where the full gravitational potential is mapped in
three dimensions. We found that the cosine between the local model and
the ``equilateral-type'' models is around 0.3-0.4 in 3D and 0.5-0.6 in 2D. That is to say, the two
distributions are quite orthogonal. 

The low cosine means that data analysis
techniques optimized for one distribution are not optimal for the
other.  Setting constraints on the local model is computationally much
simpler than for the other examples as a result of various tricks
developed in the literature \cite{Komatsu:rj}.  Our results suggest
that to fully exploit available and future data one should find ways of extending
existing techniques to apply for other 3-point function shapes.

When comparing different models of non-Gaussianity it has become
fairly standard to normalize them to have equal amplitude at
equilateral configurations. We showed that because the local model has
a fairly small signal there while the ``equilateral-type'' models peak
for these configurations, using this method to read off limits for one
model based on constraints on another can be quite misleading. For
example we found that the constraints on the ghost-inflation model are
significantly relaxed. This is mainly because when normalized at the
equilateral configurations, the ghost model is significantly less
non-Gaussian than the local model, and to a lesser extent because the data analysis
is not optimized for the ghost-inflation distribution.

\section*{Acknowledgments}
It is a pleasure to thank Nima Arkani-Hamed for useful discussions. We thank Paul Steinhardt for discussions about the ekpyrotic 
and cyclic scenarios. D.~B. and M.~Z. are supported by NSF grants AST 0098606  and 
by the David and Lucille Packard Foundation Fellowship for Science and Engineering.

\appendix
\section*{Appendix}
In this appendix we present the approximated method we used to calculate the 2d cosines and fudge factors.
We will start with the integral solution of the brightness equation \cite{Seljak:1996is},
\begin{equation}\label{line}
\frac{\Delta T}{T}(\hat{n}) = \int \frac{d^3\vec{k}}{(2\pi)^3} \zeta(\vec{k}) 
\int_0^{\tau_0} d\tau e^{i \vec{k} \cdot \hat{n} (\tau_0-\tau)} S(k,\tau) \;,
\end{equation}
where $\Delta T(\hat{n})/T$ is the fluctuation in the CMB temperature in the $\hat{n}$
direction, $S(k,\tau)$ is the CMB source function. The source function encodes the 
effects of the metric perturbations and photon fluctuations, through the integrated Sachs-Wolfe effect, 
Doppler effect, gravitational redshift, etc.., on the observed CMB. 

In the flat sky approximation one considers directions very close to some fiducial 
direction, and ignores the curvature of the sky taking 
$\hat{n}$ to lie in the plane perpendicular to the fiducial direction. This is equivalent 
to approximating the sphere in a neighborhood of a point by the tangent plane at that
point.  In this limit the equivalent of spherical harmonic transformation becomes simply a Fourier transform. 
We have 
\begin{equation}
a(\vec{l}) = \int d^2 \hat{n} \; e^{-i \vec{l} \cdot \hat{n}} \;
\frac{\Delta T}{T}(\hat{n}) \;.
\end{equation}
 
We can separate the exponential term in the line of sight integral into two 
pieces that depend of the wavevectors parallel and perpendicular to the tangent plane, 
\begin{equation}
a(\vec{l}) = \int \frac{d^3\vec{k}}{(2\pi)^3} \zeta(\vec{k}) \int_0^{\tau_0} 
\! d \tau S(k,\tau) \int d^2 \hat{n} e^{-i \vec{l} \cdot \hat{n}} 
e^{i\vec{k}^{\parallel} \cdot \hat{n} (\tau_0 - \tau)} e^{i k^{z}(\tau_0 - \tau)} \;.
\end{equation}
Evaluating the integral over $\hat{n}$ we recover a 2D $\delta$ function 
that requires $\vec{l}$ to be equal to the projected wavevector $\vec{k^{\parallel}}$ times 
the distance to the last scattering of the observed photon,         
\begin{equation}
\label{ap1}
a(\vec{l}) = \int \frac{d^3\vec{k}}{(2\pi)^3} \zeta(\vec{k}) \int_0^{\tau_0} \!d\tau S(k,\tau)  
e^{i k^{z}(\tau_0 - \tau)} (2\pi)^2 \delta^2(\vec{l}-\vec{k}^{\parallel}(\tau_0-\tau)) \;.
\end{equation}
This approximation will break down when the tangent plane needed to define a mode with wavenumber $l$ has large
deviations from the surface of the sphere that defines the last scattering surface (LSS), that is to say when considering 
large angular scales. Note that we have not assumed that recombination happens instantaneously. Although the source function is strongly peaked
around the decoupling time $\tau_R$, it has a width $\delta\tau_R$ which for our purposes cannot be ignored. 

We separate the phase $e^{ik^z(\tau_0-\tau_R)}$ in eq.~(\ref{ap1}), this factor will cancel 
when we look at n-point correlation functions of $a(\vec{l})$ because of momentum conservation. We have 
\begin{equation}
a(\vec{l}) = \int \frac{dk^z}{2\pi}e^{i k^z(\tau_0-\tau_R)} \int d^2\vec{k}^{\parallel} \zeta(\vec{k}) 
\int_0^{\tau_0} d\tau S(k,\tau) e^{i k^{z}(\tau_R - \tau)} 
\delta^2(\vec{l}-\vec{k}^{\parallel}(\tau_0-\tau)) \;.
\end{equation}
This allows us to define a radiation transfer function as,
   \begin{equation}\label{radtran}
     \Delta^T(l,k_z) = \int_0^{\tau_0} \!d\tau S(k,\tau) e^{ik^z(\tau_R-\tau)} 
        \delta^2(\vec{l}-\vec{k}^{\parallel}(\tau_0-\tau)) \;.
   \end{equation}
Comparing eq.~(\ref{radtran}) to the all sky formula \cite{Seljak:1996is}
   \begin{equation}
       \Delta^T_l(k) = \int^{\tau_0}_0 \!d\tau S(k,\tau) j_l[k(\tau_0-\tau)] \;,
   \end{equation}
we can see that spherical Bessel function encapsulates both the 3D to 2D map
of the $\delta$ function and the oscillation  present in the exponential factor
$e^{ik^z (\tau_0-\tau_R)}$.

Now we can calculate the bispectrum by taking the  ensemble average of a product of three $a(\vec{l})$,
\begin{eqnarray}
      \langle a(\vec{l_1}) a(\vec{l_2}) a(\vec{l_3}) \rangle 
      = \int d\tau_1 d\tau_2 d\tau_3 \int \frac{d^3\vec{k}_1}{2\pi} \frac{d^3\vec{k}_2}{2\pi} 
        \frac{d^3\vec{k}_3}{2\pi}  \langle \zeta(\vec{k}_1) \zeta(\vec{k}_2) \zeta(\vec{k}_3) \rangle       
       S(k_1,\tau_1) S(k_2,\tau_2) S(k_3,\tau_3) \nonumber 
      \\ \times  e^{i(\tau_R-\tau_1) k_1^z} 
       e^{i(\tau_R-\tau_2)k_2^z} e^{i(\tau_R-\tau_3)k_3^z} 
       \delta^2(\vec{l}_1-\vec{k}^{\parallel}_1(\tau_0-\tau_1))
       \delta^2(\vec{l}_2-\vec{k}^{\parallel}_2(\tau_0-\tau_2)) 
       \delta^2(\vec{l}_3-\vec{k}^{\parallel}_3(\tau_0-\tau_3)) \;.
\end{eqnarray}
Now we use $\langle \zeta(\vec{k}_1)\zeta(\vec{k}_2)\zeta(\vec{k}_3) 
 \rangle = (2\pi)^3 \delta^3(\vec{k}_{123}) F(k_1,k_2,k_3)$ and assume that 
 $\delta \vec{k}^{\parallel} \cdot \vec{\nabla}_{k} F(k_1,k_2,k_3)$ is small, 
  where $\delta k^{\parallel}$ is the variation in $k^{\parallel}$ for a given $l$ as the tangent plane sweeps across the width of the last scattering 
surface. It is clear from geometry that $\delta k^{\parallel}/k^{\parallel}$ 
will be order $\delta \tau_R / \tau_R \sim 10^{-2}$. The 3-point functions we are considering 
do not have sharp features so this assumption will allow us to use an average 
$k^{\parallel}$ in our evaluation of the primordial bispectrum without introducing a large error. 
This is equivalent to interchanging the line of sight integral and the integral over Fourier space 
and evaluating $\vec{k}^{\parallel}$ at $\vec{l}/(\tau_0-\tau_R)$,
   \begin{equation}\label{ap2}
      \langle a(\vec{l}_1)a(\vec{l}_2)a(\vec{l}_3) \rangle = (\tau_0 - \tau_R)^2\delta^{(2)}(\vec{l}_{123})
       \!\int \! dk^z_1 dk^z_2 dk^z_3 \delta (k^z_{123}) F(k_1',k_2',k_3') 
       \tilde{\Delta}^T(l_1,k_1^z) \tilde{\Delta}^T(l_2,k_2^z) 
       \tilde{\Delta}^T(l_3,k_3^z) 
   \end{equation}   
   where $k'$ means $k$ evaluated such that $\vec{k}^{\parallel} = \vec{l}/(\tau_0-\tau_R)$ and 
   \begin{equation}
        \tilde{\Delta}^T(l,k^z) = \int_0^{\tau_0} \frac{d\tau}{(\tau_0-\tau)^2} 
          S(\sqrt{(k^z)^2 + l^2/(\tau_0-\tau)^2},\tau) 
          e^{ik^z(\tau_R-\tau)} \;. 
   \end{equation}

Using the definition for the bispectrum,  $\langle a(\vec{l}_1)a(\vec{l}_2)a(\vec{l}_3) \rangle 
   = (2\pi)^2\delta^{(2)}(\vec{l}_{123}) B(l_1,l_2,l_3)$ we get
\begin{equation} \label{bispectflat}
B(l_1,l_2,l_3) = \frac{(\tau_0 - \tau_R)^2}{(2\pi)^2} \int
dk^z_1dk^z_2dk^z_3 \delta^{(1)}(k^z_{123}) F(k'_1,k'_2,k'_3)  
\tilde{\Delta}^T(l_1,k^z_1) \tilde{\Delta}^T(l_2,k^z_2) \tilde{\Delta}^T(l_3,k^z_3) \;.
\end{equation}

This formula is completely generic, it can be used to calculate the flat sky bispectrum produced by 
any primordial perturbation field. It is much easier to handle numerically and was the basis of our evaluation 
of the dot products presented in Table \ref{results}.

In the flat sky limit we have a 2D $\delta$ function which forces the 2D modes to form a closed triangle.
In the full sky calculations this is enforced by the Wigner $3j$ symbols of eq.~(\ref{gaunt}).

\end{document}